%% file: ms.tex
\documentclass{article}

\usepackage{arxiv}

\usepackage[utf8]{inputenc} 
\usepackage[T1]{fontenc}    
\usepackage{hyperref}       
\usepackage{url}            
\usepackage{booktabs}       
\usepackage{amsfonts}       
\usepackage{nicefrac}       
\usepackage{microtype}      
\usepackage{multicol,caption}
\usepackage{graphicx}
\usepackage{long2}

\usepackage{amssymb}
\usepackage{latexsym}

\usepackage{xcolor}
\usepackage{longtable}
\usepackage{supertabular}
\usepackage{cuted}
\usepackage[round]{natbib}

\usepackage{ltablex}
    \newcolumntype{L}{>{\raggedright\arraybackslash}X}
    \newcolumntype{S}{>{\hsize=.35\hsize\raggedright\arraybackslash}X}
	\newcolumntype{m}{>{\hsize=.3\hsize}X}

\title{A survey on shape-constraint deep learning for medical image segmentation}

\author{
  Simon Bohlender \\
  Department of Computer Science\\
  TU Darmstadt\\
  Darmstadt, Germany \\
   \And
 Ilkay Oksuz \\
  Computer Engineering Department\\
  Istanbul Technical University\\
  Istanbul, Turkey \\\\
  School of Biomedical Engineering Imaging Sciences\\
  King\textquotesingle s College \\
  London, U.K.\\
   \And
 Anirban Mukhopadhyay \\
  Department of Computer Science\\
  TU Darmstadt\\
  Darmstadt, Germany \\
}

\newenvironment{Figure}
  {\par\medskip\noindent\minipage{\linewidth}}
  {\endminipage\par\medskip}
\begin{document}

\maketitle
\begin{abstract}
Since the advent of U-Net, fully convolutional deep neural networks and its many variants have completely changed the modern landscape of deep learning based medical image segmentation. However, the over dependence of these methods on pixel level classification and regression has been identified early on as a problem. Especially when trained on medical databases with sparse available annotation, these methods are prone to generate segmentation artifacts such as fragmented structures, topological inconsistencies and islands of pixel. These artefacts are especially problematic in medical imaging since segmentation is almost always a pre-processing step for some downstream evaluation. The range of possible downstream evaluations is rather big, for example surgical planning, visualization, shape analysis, prognosis, treatment planning etc. However, one common thread across all these downstream tasks is the demand of anatomical consistency. To ensure the segmentation result is anatomically consistent, approaches based on Markov/ Conditional Random Fields, Statistical Shape Models are becoming increasingly popular over the past 5 years. In this review paper, a broad overview of recent literature on bringing anatomical constraints for medical image segmentation  is given, the shortcomings and opportunities of the proposed methods are thoroughly discussed and potential future work is elaborated. We review the most relevant papers published until the submission date. For quick access, important details such as the underlying method, datasets and performance are tabulated.
\end{abstract}

\begin{multicols}{2}
\input{sections/introduction}
\end{multicols}
\input{sections/relatedwork}
\begin{multicols}{2}
\input{sections/discussion}
\input{sections/conclusion}

\bibliographystyle{plainnat}
\bibliography{refs,ilkay}

\end{multicols}

\end{document}

%% file: sections/introduction.tex
\section{Introduction}\label{intro_main}



Semantic segmentation is the task of predicting the category of individual pixels in the image which has been one of the key problems in the field of image understanding and computer vision for a long time. It has a vast range of applications
such as autonomous driving (detecting road signs, pedestrians and other road users), land use and land cover classification, image search engines, medical field (detecting and localizing the surgical instruments, describing the brain tumors, identifying organs in different image modalities). This problem has been tackled by a combination of machine learning and computer vision, approaches in the past. Despite their popularity and success, deep learning era changed main trends. Many of the problems in computer vision - semantic segmentation among them - have been solved with convolutional neural networks (CNNs) .

Incorporating prior knowledge into traditional image segmentation algorithms has proven useful for obtaining more accurate and plausible results. The highly constrained nature of anatomical objects can be well captured with learning based techniques. However, in most recent and promising techniques such as CNN based segmentation it is not obvious how to incorporate such prior knowledge. 
Segmenting images that suffer from low-quality and low signal-to-noise ratio without any shape constraint remains problematic even for CNNs. Though it has been shown that incorporation of shape prior information significantly improves the performance of the segmentation algorithms,
incorporation of such prior knowledge is a tricky practical problem. In this work, we provide an overview of efforts of shape prior usage in deep learning frameworks.

\subsection{Yet another review paper}

There already appeared a variety of review papers about shape modelling and deep learning for medical image segmentation in the recent past. \cite{DBLP:journals/mia/McInerneyT96} presents various approaches that apply deformable models. \cite{DBLP:journals/pr/PengZZ13} deals with different categories of graph-based models where meaningful objects are represented by sub-graphs. The review by \cite{DBLP:journals/mia/HeimannM09} is about statistical shape models and concentrates especially on landmark-based shape representations. \cite{Elnakib2011} also reviews different shape feature based models, that include statistical shape models, as well as deformable models. A more recent review by \cite{DBLP:journals/corr/NosratiH16} provides insights into segmentation models that incorporate shape information as prior knowledge. Later surveys of \cite{DBLP:journals/mia/LitjensKBSCGLGS17}, \cite{DBLP:journals/corr/RazzakNZ17}, \cite{RIZWANIHAQUE2020100297} and \cite{DBLP:journals/corr/abs-2009-13120} shift their focus to deep learning approaches. \cite{DBLP:journals/jdi/HesamianJHK19} and \cite{DBLP:journals/corr/abs-1910-07655} present different network architectures and training techniques, whereas \cite{DBLP:journals/corr/abs-2011-08018} take it a step further and reviews prior-based loss functions in neural networks.

Since deep learning became the method of choice for many computer vision tasks, including medical image segmentation, we focus our review on models that combine neural networks with explicit shape models in order to incorporate shape knowledge into the segmentation process. Segmentation models solely based on neural networks usually do not incorporate any form of shape knowledge. They are based on traditional loss functions that only regard objects at pixel level and do not evaluate global structures. The papers we present in this review improve these networks by combining them with additional models that are especially built with shape in mind. This is also the point that delimits this review from existing surveys which either focus mostly deep learning approaches or on traditional shape and deformable model methods, but not on the combination of both.

The explicit models applied in this review can be divided into three main categories as shown in Figure \ref{rel_work}: 1) Conditional or Markov Field models that establish connections between different pixel regions 2) Active/Statistical Shape Models that learn a special representation for valid shapes 3) Active Contour Models or snakes that use deformable splines for shape detection. These models are either applied as pre-processing steps to create initial segmentations, post-processing steps to refine the neural network segmentations, or used in multi-step models consisting of various models along a specific pipeline. 

We are aware that the field is heavily shifting from explicit ways of modeling shape to more implicit approaches where networks are trained in an end-to-end way.Up and coming Works  propose more intelligent loss functions that no longer require additional explicit shape modelling, but only consist of a single neural network. \cite{DBLP:journals/corr/abs-2009-13755} proposed a new geometric loss for lesion segmentation. Other examples are \cite{DBLP:journals/cars/MohagheghiF20} and \cite{DBLP:journals/access/HanKKC20} where the loss contains shape priors. \cite{DBLP:conf/miccai/LiWSZ20} introduces a spatially encoded loss with a special shape attention mechanism. \cite{DBLP:journals/corr/abs-1910-01877} uses a topology based loss function. 

However the overwhelming majority of articles combine neural networks and explicit models to introduce shape knowledge.
This combination often stems from a rather principled engineering design choice (as shown in Figure~\ref{rel_work}) which is not detailed in any of the previous review articles.
This review focuses on this overarching design principle of shape constraint which, along with being a quick access guide to explicit approaches, will work as a research catalyzer of implicit constraints.  

%% file: sections/relatedwork.tex
\begin{multicols}{2}
\begin{Figure}
    \centering
    \includegraphics[scale=.45]{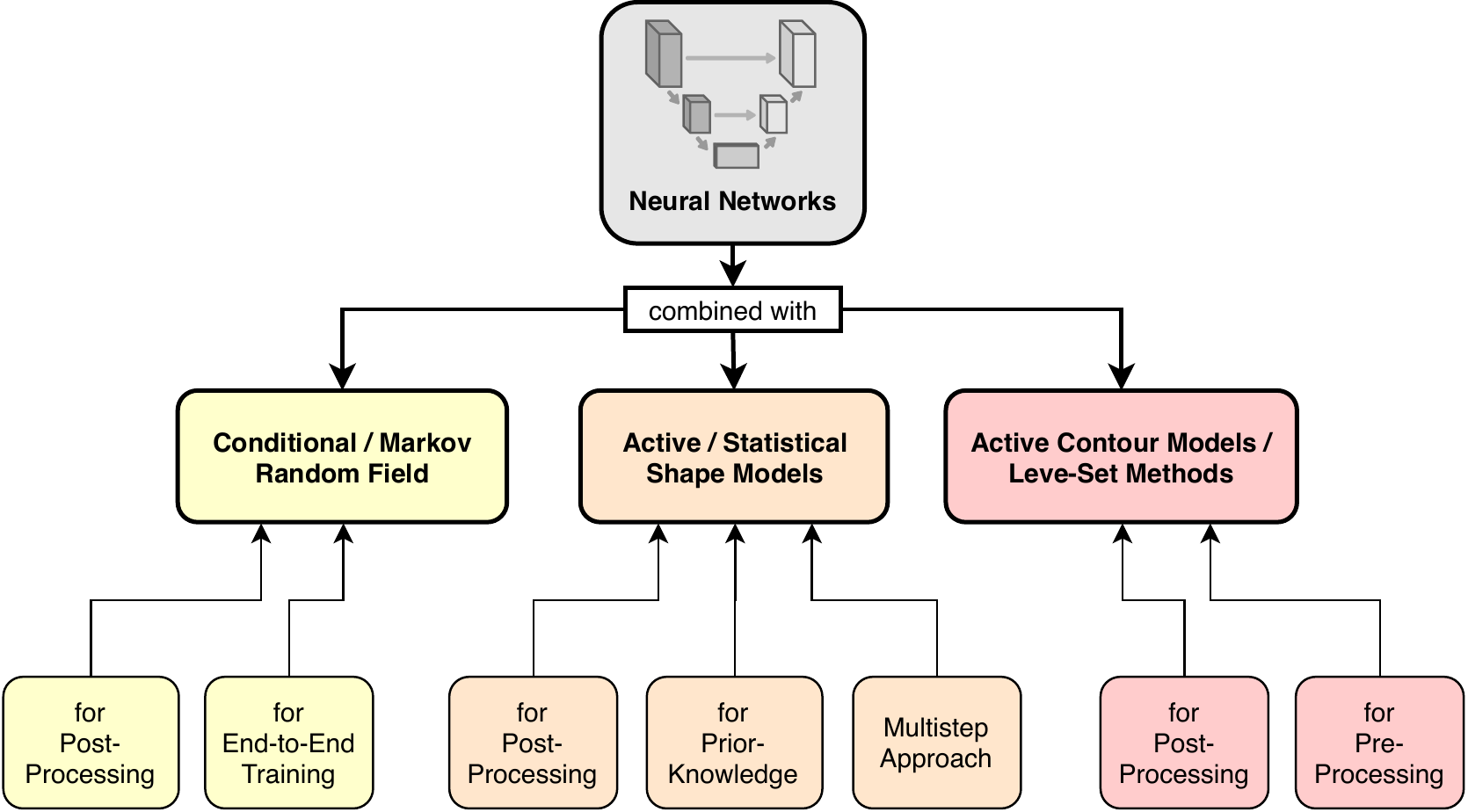}
	\captionof{figure}{Overview of related work approaches}
	\label{rel_work}
\end{Figure}

\section{CRF / MRF approaches}

Markov Random Fields (MRF) \cite{DBLP:conf/eccv/Li94a} belong to the domain of graphical models and model relationships between pixels or high-level features with a neighborhood system. The label probability of a single pixel is thereby conditioned on all neighboring pixels which allow to model contextual constraints. The maximum \textit{a posteriori} probability (MAP) can then be calculated by applying the Bayes rule.
Conditional Random Fields (CRF) \cite{DBLP:conf/icml/LaffertyMP01} are an extension of MRFs and allow to incorporate arbitrary global features over regions of pixels. For medical image segmentation this means that they generate smooth edges by using this global knowledge about surrounding regions which is a reason why the are often applied alongside neural networks to perform medical image segmentation.
 
\paragraph{CRFs used for postprocessing}
The largest category of methods that utilize CRFs or MRFs apply them as a pos-tprocessing step. A large portion of papers focus on the straight-forward approach where the CNN generates initial segmentations maps which are directly passed to a CRF or MRF model as inputs for further refinements. These approaches are evaluated on a variety of anatomies and mostly differ in the utilized network architectures but follow the same idea. They are applied to lung nodules (\cite{DBLP:conf/micad/YaguchiATO19}, \cite{DBLP:conf/isbi/GaoXLWNSM16}), retinal vessel (\cite{DBLP:conf/isbi/FuXWL16}), brain tumor (\cite{DBLP:conf/miccai/ZhaoWSLFZ16}, \cite{journals/johe/zeju2017}), cervical nuclei (\cite{DBLP:journals/access/LiuZSLZG18}), eye sclera (\cite{DBLP:conf/icb/MesbahMM17}), melanoma (\cite{DBLP:conf/icpr/LuoY18}), ocular structure (\cite{DBLP:conf/miccai/NguyenPMSPHWCS18}), left atrial appendage (\cite{DBLP:journals/titb/JinFWYLLZ18}), lymph node (\cite{DBLP:conf/miccai/NoguesLWRBLSTS16}), liver (\cite{DBLP:conf/miccai/Dou0JYQH16}) and prostate cancer lesion (\cite{DBLP:conf/isbi/CaoZSBMERS19}) segmentation tasks. A slightly different approach for skin lesion detection by \cite{DBLP:journals/access/QiuCQZ20} is based on the same idea, but uses not just a single CNN network, but an ensemble of seven or fifteen which are combined inside the CRF. Two other approaches to highlight here for brain region (\cite{DBLP:journals/jms/ZhaiL19}) and optical discs in fundus image (\cite{8986563}) segmentation integrate a special attention mechanism into their networks with the motivation to improve the segmentations by detecting and exploiting salient deep features. Another special version that operates on weakly segmented bounding box images for fetal brain \& lung segmentation is introduced by \cite{DBLP:journals/tmi/RajchlLOKPBDRHK17}. Given the initial weak segmentations, the model iteratively optimizes the pixel predictions with a CNN followed by a CRF to obtain the final segmentation maps.\\
Instead of CRFs, \cite{DBLP:conf/isbi/ShakenTFLKPK16} use a MRF to impose volumetric homogenity on the outputs of a CNN for subcortical region segmentation. MRFs are also utilized in the approach shown by \cite{DBLP:journals/jms/XiaYZ19} for kidney segmentation where the MRF is integrated into a SIFT-Flow model. \\
Besides these classical approaches, another method that came up focused on cascading CNN networks that generate segmentations in a coarse-to-fine fashion. \cite{DBLP:journals/neuroimage/WachingerRK18} use this strategy with a first network that segments fore- from background pixels in brain MRIs and a second one that classifies the actual brain regions. The same method is also used by \cite{shen2017} for brain tumor segmentation, by \cite{DBLP:journals/mia/DouYCJYQH17} for liver and whole heart segmentation, and by \cite{DBLP:conf/miccai/ChristEETBBRAHD16} for liver-based lesion segmentation.\\
A somewhat different cascading structure, for brain tumor segmentation, is introduced by \cite{DBLP:journals/access/HuGZDXHCG19} where multiple subsequent CNNs are used to extract more discriminative multi-scale features and to capture dependencies. \cite{DBLP:journals/access/FengGQ20} extend this version on the task of brain tumor segmentation with the introduction of residual connections that improve the overall performance.
Similar to the cascading methods, there are CNNs with two pathways that combine two parallel networks on different resolution levels that aim for capturing larger 3D contexts. The approach was originally introduced by \cite{DBLP:conf/miccai/AlansaryKDKRMRH16} for placenta segmentation, but was also applied in \cite{DBLP:conf/miccai/CaiLXXY17} to the task of pancreas segmentation. \cite{DBLP:journals/mia/KamnitsasLNSKMR17} proposes another related approach where two parallel networks, a FCN that extracts a rough mask and a HED that outputs a contour, are fused inside a CRF. In the approach by \cite{DBLP:conf/icmip/ShenDLCQ18} that deals with brain tumor segmentation, a third path is added where in total three concurrent FCNs are trained based on different filtered (gaussian, mean, median) input images. After each network an individual CRF is applied and their results are fused in a linear regression model. \\

\begin{Figure}
	\centering
    \includegraphics[scale=.55]{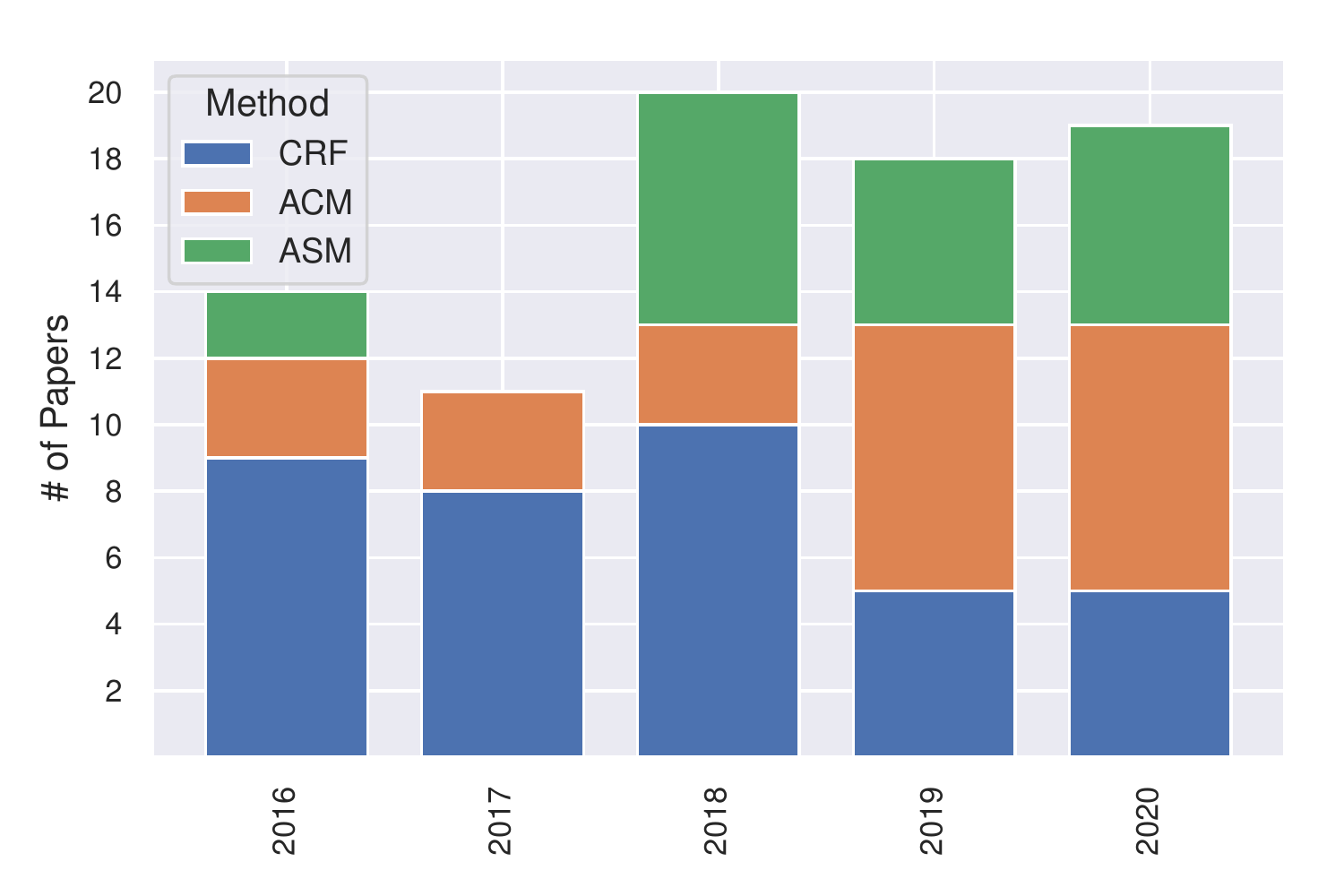}
    \captionof{figure}{Overview of relevant papers per year for each category}
\end{Figure}

\paragraph{Training CNN and CRF models end-to-end}
The idea of integrating CRF models directly into neural networks origins from the task of semantic image segmentation and was introduced by \cite{DBLP:conf/iccv/0001JRVSDHT15}. They combine the strengths of both models into a unified framework that allows end-to-end training. Broken down, the basic task of CRFs is to minimize an energy term with an iterative mean field approximation. Since CRFs are graphical models, each iteration step can be formulated as a stack of CNN layers. Multiple iterations can then be implemented by repeatedly executing this stack or alternatively as an equivalent Recurrent Neural Network (RNN). The resulting network is denoted as a CRF-RNN and can be applied on top of any CNN architecture.
\cite{DBLP:conf/miccai/FuXLW016} are the first to transfer this method to medical image segmentations with a model called \textit{DeepVessel} for the task of retinal vessel segmentation. For the same task \cite{10.1007/978-981-10-5230-9_17} achieve similar results by using a slightly deeper base CNN network with more convolution layers. Besides retinal vessel, CRF-RNN approaches are applied to a variety of other anatomical structures. \cite{DBLP:conf/miccai/ZhaoWSLFZ16} applies them to brain tumor segmentation and extend it with some additional pre- and post-processing steps later on \cite{DBLP:journals/mia/ZhaoWSLZF18}. \cite{DBLP:journals/cars/XuZL18} uses a V-Net combined with CRF-RNN for bladder segmentation and in \cite{DBLP:journals/corr/abs-1807-07464} they are also applied on brain tumor as well as prostate segmentation with 3D multi-modal images. Analogues \cite{DBLP:journals/corr/abs-1811-03549} utilizes a U-Net as their base-network to deal with white matter lesion segmentation. On the same idea as CRF-RNN \cite{DBLP:journals/access/DengSWZN20} uses a CRF-Recurrent Regression based Neural Network (CRF-RRNN) integrated with a heterogeneous CNN for brain tumor segmentation where the combined network can also be trained end-to-end. Instead of using a full RNN,  \cite{DBLP:journals/corr/abs-2008-04488} propose a method where MRF is integrated into the segmentation network as a block of local and global convolution layers that take the CNN output as unary potentials to calculate the corresponding pairwise potentials.
\end{multicols}

\input{tables/crf}

\begin{multicols}{2}
\section{Shape model based approaches}
The second category of model assumptions often combined with CNNs are active shape models (ASM) \cite{DBLP:journals/cviu/CootesTCG95} or probabilistic active shape models (PASM). ASMs require a training set with a fixed number of manually annotated landmark points of the segmented object. Each point represents a particular part of the object and has to be in the same position over all images. These annotated shapes are then iteratively matched and a mean shape is derived. The landmark points show different variabilities that are modeled by a Point Distribution Model (PDM). Performing a principal component analysis (PCA) and weighting the eigenvectors allows creating new shapes in the allowed variability range. For detecting an object in an unknown image an algorithm is used that updates pose and shape parameters iteratively to improve the match until convergence. An extension to this approach are probabilistic ASMs (PASM) \cite{DBLP:conf/miccai/WimmerSH09}. They impose a weaker constraint on shapes which allows more flexible contours with more variations from the mean shape. This is achieved by introducing a probabilistic energy function which is minimized in order to fit a shape to a given image. The model's ability to generalize is thereby improved and the segmentation results outperform standard ASMs. 

\paragraph{Shape Models for post-processing}
Though CNN based segmentation models yield good segmentation results, they tend to produce anatomically implausible segmentation maps that can contain detached islands or holes at parts where they do not occur in reality. Since shape models represent valid and anatomically plausible shapes, it makes sense to apply them in post-processing steps to regularize initial CNN segmentations and transform them into a valid shape domain. 
\cite{DBLP:journals/tmi/XingXY16} take up this idea and apply it to nucleus segmentation. The initial segmentations are generated by a CNN and the post-processing step includes a sparse selection-based shape model for top-down shape inference, which is more insensitive to object occlusions compared to PCA-based shape models, and an additional deformable model for bottom-up shape deformation. Also \cite{DBLP:journals/access/Hsu19} follows this strategy for segmentation and tracking of the left ventricle. They swap out the CNN for a Faster-RCNN and use an improved ASM that allows to obtain matching points in greater ranges. \cite{DBLP:journals/cars/FauserSBHKKSM19} continue on improving the ASM by using a probabilistic ASM that is more flexible and allows leaving the shape space. The segmentation of the left ventricle is performed by combining the results of three CNN-PASM models for each dimension. Another modified ASM is proposed by \cite{DBLP:journals/tip/MedleySN20}. The authors use Expectation-Maximization to deal with outliers during optimizing the ASM. They also evaluate different ASM features and conclude that a CNN that learns the input feature maps for the EM-ASM performs best. Besides improving on the ASM a different approach by \cite{DBLP:journals/mia/KarimiZMAMSAS19} aims for generating better predictions with an ensemble of U-Net like CNN models with different filters and parameters. In their approach a SSM model, based on the thresholded segmentations from all individual models, is only applied if the disagreement between the ensemble models becomes to high. Instead of using the CNN for generating segmentation maps, it is also sufficient to only predict bounding boxes as initializations for ASMs. Such an approach is applied by \cite{DBLP:conf/isbi/TabriziMCJL18} on kidney segmentation where a fuzzy-ASM produces the final segmentations. \cite{DBLP:journals/tmi/LiHTCST18} also uses a CNN for bounding box prediction, but adds an intermediate step before utilizing a statistical shape model for myocardial segmentation, in which a random forest classifier builds probability maps from the given bounding boxes. Another tree model, more specific an adaptive feature learning probability boosting tree (AFL-PBT) is also utilized by \cite{DBLP:journals/access/HeXHJ18} as an initial step to classify voxels for prostate segmentation. A subsequent CNN then extracts boundary probability maps and a three-level ASM is employed to generate final segmentations. 

\paragraph{Shape Models for prior knowledge}
In this second paragraph we present some papers where the shape models are applied pre-hoc before any deep learning network. Two straight forward models for this category are proposed by \cite{DBLP:conf/miip/ChengRLWTGMACSM16} and \cite{DBLP:conf/miip/FanZWND20}. In \cite{DBLP:conf/miip/FanZWND20} a 3D U-Net-like CNN segments Itra-Cholear anatomy based on initial segmentations from an ASM and the original CT images. \cite{DBLP:conf/miip/ChengRLWTGMACSM16} on the other hand use a CNN for refining initial segmentations from an Active Appearance Model (AAM) that produces only coarse prostate segmentations. The AAM is basically an extended shape model that adds an additional texture model for better fitting capabilities. The other two models already introduce some pipeline-like approaches, but use both a shape model as prior knowledge. The pipeline for subcortical region segmentation in \cite{DBLP:journals/corr/abs-1802-01268} starts with a pre-processing SVM that classifies sagittal slices into groups of similar shape. The prior ASM then creates rough segmentations for each group which are finalized by a CNN. Further the authors propose an optional CRF model for post-processing.  \cite{DBLP:conf/midl/NguyenPHWRSSC19} introduce the ASM as a more traditional prior for uveal melanoma segmentation where it is used as a constraining term for a CRF model that is based on Grad-CAM (class activation maps) heatmaps. The final segmentations are again generated with a U-Net that combines the CRF with original input CTs.

\begin{Figure}
	\centering
    \includegraphics[scale=.55]{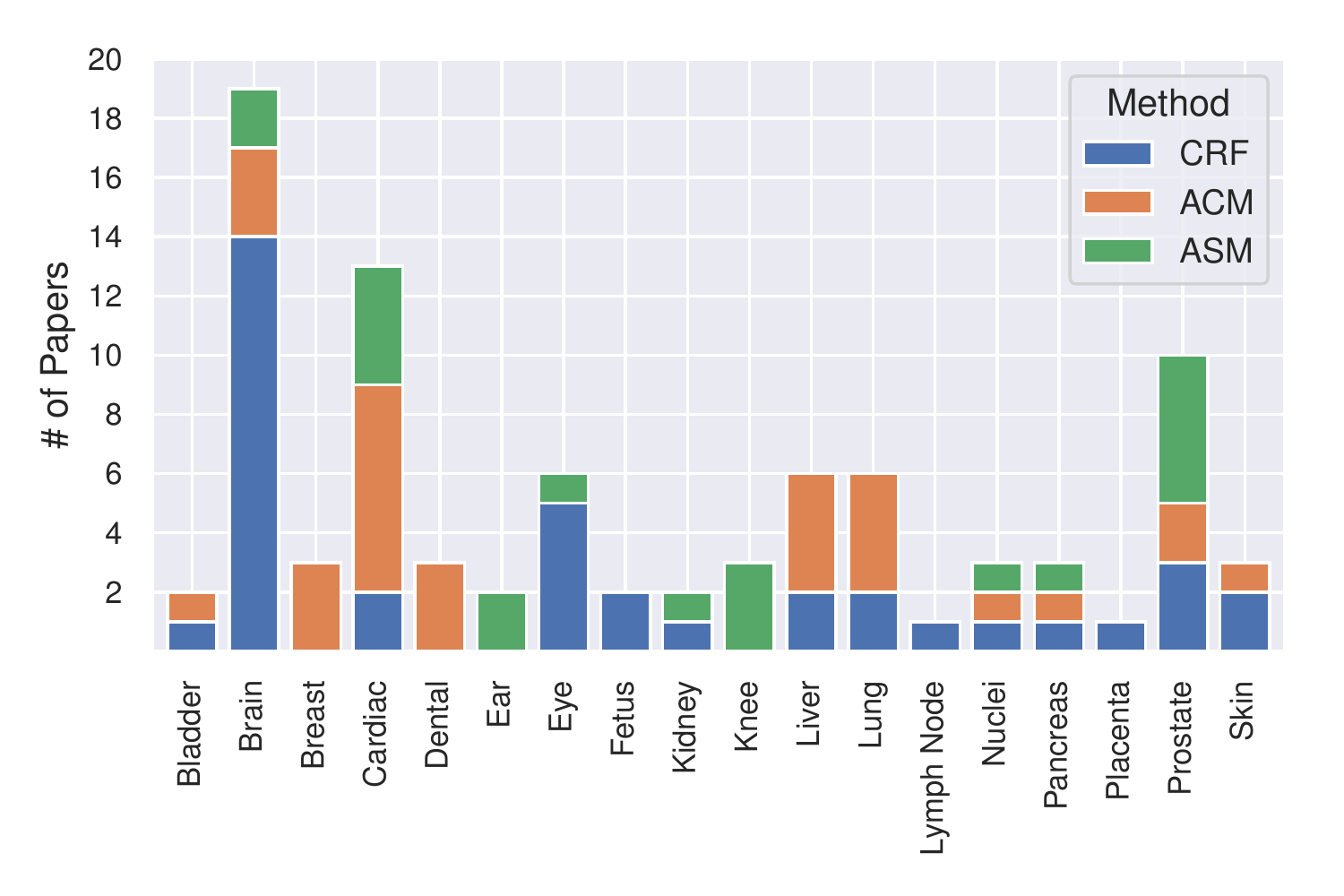}
    \captionof{figure}{Overview of anatomical structures examined in the relevant papers}
\end{Figure}

\paragraph{Pipeline approaches with multiple CNN and ASM models}
The last category for combining shape models and neural networks contains all approaches that consist of different models arranged along pipelines. The motivation is to process input images stage-wise or in a coarse-to-fine way that allows to capture more information and hence result in more accurate segmentation maps. In the models by \cite{jorunal/oac/tack2018} for knee menisci, \cite{DBLP:journals/mia/AmbellanTEZ19} for knee bone \& cartilage, and \cite{10.3389/fnins.2020.00015} for hippocampus segmentation, the pipelines combine multiple CNNs and SSMs. All three start with initial 2D U-Nets regularized by SSMs which are used to extract smaller 3D subvolumes. \cite{jorunal/oac/tack2018} and \cite{DBLP:journals/mia/AmbellanTEZ19} apply an additional 3D U-Net afterwards, whereas \cite{10.3389/fnins.2020.00015} uses three U-Nets and averages their predictions to obtain final segmentations. \cite{DBLP:journals/mia/AmbellanTEZ19} further continues after this step and utilizes a second 3D SSM model to obtain the knee bone segmentations and even applies a third U-Net to segment the cartilage afterwards.
Besides these typical pipelines, there are also some hybrid approaches we count to this category that integrate shape models and neural networks. They use special CNNs that directly predicts the parameters of an SSM, which are the shape coefficients (weights for the modes of variations), the pose parameters.  \cite{DBLP:journals/corr/abs-2007-11726} use such a SSM-Net inside a small pipeline for prostate segmentation. They propose an inception-based network that directly predicts parameters of the SSM which can be back-translated into a prostate contour prediction. Parallel to this, a residual U-Net generates probability maps from the inputs. The final segmentations are generated by averaging the outputs of both models. The method of \cite{DBLP:journals/corr/abs-2010-08952} for left ventricle segmentation is based on the same idea, but removes the small pipeline. Instead they modify the CNN and add a third output which is an actual distance map. A special loss function is used to train the network toward optimizing the segmentation map alongside the SSM parameters. A nearly identical approach by \cite{DBLP:journals/cars/KarimiSKNS18} is applied to prostate segmentation. Their CNN predicts center position of the prostate, the shape model parameters, and a rotation vector which are passed to a final layer that outputs the coordinates of the landmark points which resemble the a final segmentation map. \cite{DBLP:conf/miccai/SchockKAHKTCAKN20} relies on the same method for knee bone \& cartilage segmentation, but extend it with additional pre- and post-processing steps. They add a preprocessing 2D U-Net that detects initial bone positions and crop the volume into subvolumes which only contain the femur or tibia bone. Afterwards their SSM-Net comes into place that predicts the SSM parameters and the actual landmarks in a subsequent PCA layer. An additional fine-tuning step then generates the cartilage segmentations with a 3D U-Net based on subvolumes centered at the bones' landmark points. Rather than integrating the SSM and CNN, \cite{DBLP:conf/miccai/MaLWE18} introduces a Bayesian model that integrates both, the CNN and a robust kernel SSM (RKSSM) for the task of pancreas segmentation. At first the RKSSM is initialized to fit the detected ROI of a Dense U-Net. A Gaussian Mixture Model afterwards guides the shape adaption and iteratively projects the adapted shape onto the RKSSM until convergence which results in the final segmentation map.
\end{multicols}

\input{tables/asm}

\begin{multicols}{2}
\section{Active contour approaches}
A last type of models that often combined with deep learning models to incorporate shape knowledge are \textit{Active Contour Models} (ACM) \cite{DBLP:journals/ijcv/KassWT88} , also known as \textit{snakes}. A snake is a deformable controlled continuity spline that is pushed towards edges or contours by minimizing an energy function under the influence of different forces and constraints. It consists of an internal energy that keeps the contour continuous and smooth, an image energy that attracts it to contours, and an external constraint force that adds user-imposed guidance. A similar approach are \textit{level set functions} (LSF) introduced by \cite{DBLP:journals/robotica/Andrew00a} and firstly applied to image segmentation by \cite{DBLP:journals/pami/MalladiSV95}. An LSF is a higher dimensional function where a contour is defined as its zero level set. With a \textit{speed function}, derived from the image, that controls the evolution of the surface over time, a Hamilton-Jacobi partial differential equation can be obtained.

\paragraph{ACM models for post-processing}
Since ACM models are based on the idea of evolving a contour, it makes sense to apply them as a post-processing step to improve an initial segmentation map.
An early model by \cite{jorunal/mep/middelton2004} uses only a simple multilayer perceptron (MLP) that creates binary pixel-wise boundary predictions for lung segmentation. Since these are very rough and contain misclassifications the ASM is used to improve and close the contour. \cite{DBLP:journals/sivp/SalimiPM18} is also based on an MLP, but adds an vector field convolution to the ACM to make it more robust for prostate segmentation.\\
However, the more recent ACM post-processing models are exclusively based on different CNN architectures and are applied to a variety of anatomies. \cite{DBLP:conf/bibm/LiLYW17} use a FCN that is refined by a classic ACM for left ventricle segmentation. The same approach is taken by \cite{jorunal/mp/guo2019} for liver segmentation and \cite{DBLP:conf/iconip/ZhaoWFQ18} utilize it for nucleus segmentation. In the approaches by \cite{journal/jomi/xu2019} the ACM refinements are not yet the final steps and additional adaptive ellipse fitting is used to segment breast nuclei. \cite{journal/mp/hu2018} and \cite{joruanl/jomiahi/fang2019} transfer the basic refinement method to breast tumor detection with a phase-based ACM that improves over multiple iterations. A different slightly modified ACM post-processing method is based on geodesic computations and is further used by \cite{DBLP:conf/miip/MaY19} for dental root segmentation and \cite{DBLP:conf/ijcnn/NunesMSBF20} for lung segmentation. \cite{DBLP:journals/sivp/ZhangWLW20} also introduces a special ACM that integrates a fourth-order partial differential equation and segments plaque based on an initial R-CNN segmentation. Instead of just refining an initial CNN predicted per-pixel segmentation map, \cite{DBLP:journals/mbec/SilvaDFFSPC20} use a Chan-Vese ACM to generate prostate segmentation on DCNN coarsely classified superpixels which only represent rough initialization for the contour model. The authors of \cite{DBLP:conf/ijcnn/KotKSC20} further separate the two models where the CNN masks bone tissue which is removed for the ACM to segment brain tumors. The last special approach in the ACM category by \cite{DBLP:conf/miccai/ZhangDL20} is a hybrid model that integrates an ACM into a U-Net. The resulting deep active contour network (DACN) is end-to-end trainable with a special ACM based loss function and automatically segments cervical cells and skin lesions.
Besides ACM, another large number of approaches rely on level set functions (LSF). Same as before a CNN is used for generating initial segmentation maps which are then refined by the LSF. \cite{DBLP:journals/corr/abs-1908-06933} uses this for brain, liver, lung segmentation, \cite{DBLP:conf/isicdm/GongZZZG19} for pancreas segmentation, \cite{10.1117/12.2556928} for ventricle and liver segmentation, and \cite{DBLP:journals/cbm/XieSC20} for left ventricle segmentation. Some extra processing is made in \cite{YANG2021108} for dental pulp segmentation where the initial CNN segmentations are used to calculate elliptic curves which are used to guide the LSF. In general, for the LSF it is often sufficient to initialize them only with a rough bounding boxes or region of interest annotations. So, \cite{DBLP:conf/cinc/0003WLZ19} use a Faster RCNN to generate location boxes of left atriums which serve as input for the LSF after Otsu thresholding. \cite{DBLP:journals/mia/AvendiKJ16} inserts an additional step between CNN ROI detection and LSF segmentation where the initial left-ventricle shape is inferred with an stacked auto-encoder. In comparison to these two approaches, in \cite{journal/mp/chu2016} the CNN is not used to predict ROI, but to classify if an ROI is part of the bladder. The outputs are then refined by three different 3D LSF and a final 2D LSF afterwards.
Another idea is to use recurrent pipelines where the segmentations are refined iteratively. Such an approach is introduced by \cite{DBLP:conf/miccai/TangVZCJ17} where both models are integrated into a FCN-LSF. The method is used for left ventricle and liver segmentation with semi-supervised training where the LSF gradually refines the segmentation and backpropagates a loss to improve the FCN. \cite{DBLP:journals/tmi/HoogiSVR17} proposed a different iterative process. Hereby the CNN estimates if the zero level set is inside, outside or near the lesion boundary. Based on these the LSF parameters are calculated and the contour is evolved. The process then repeats until convergence. 

\paragraph{Using a CNN to refine ACM segmentations}
Besides the majority of approaches that use ACMs for post-processing, there are also methods where ACMs are used to obtain the initial segmentations or are guided by CNNs. The earliest of these approaches by \cite{DBLP:conf/socpar/AhmedMK09} uses an ACM to remove skull tissue from images and applies a simple artificial neural network to classify the remaining brain regions. \cite{DBLP:journals/corr/RupprechtHBN16} introduce an approach where the ACM is guided by the CNN. The ACM generated rough segmentations of the left ventricle. A CNN then predicts vectors on patches around each pixel of this initial contour that point towards closes object boundary points and are used to further evolve the contour. The latest method for this category by \cite{DBLP:journals/eswa/KasinathanJGRFP19} also uses the ACM to generate initial segmentations, more specific it segments all lung nodules. A post-processing CNN afterwards classifies them or removes false positives.
\end{multicols}

\input{tables/contours}

\begin{multicols}{2}
\section{Topology based Approaches}

An alternative approach to integrating shape priors into network-based segmentation was presented in \cite{Chung2019}.
Here, the segmentation started with a candidate shape which was topologically correct (and approximately correct in terms of its shape), and the network was trained to provide the appropriate deformation to this shape such that it maximally overlapped with the ground truth segmentation.

Such methods can be considered to have a `hard prior' rather than the `soft-prior' of the methods presented above in the sense that the end result can be guaranteed to have the correct shape.
However, this approach may be limited by a requirement that the initial candidate shape be very close to an acceptable answer such that only small shape deformations are needed.
A further potential issue is that the deformation field provided by the network may need to be restricted to prevent the shape from overlapping itself and consequently changing its topology.

The differentiable properties of persistent homology \cite{Edelsbrunner2000} make it a promising candidate for the integration of topological information into the training of neural networks.
The key idea is that it measures the presence of topological features as some threshold or length scale changes.
Persistent features are those which exist for a wide range of filtration values, and this persistence is differentiable with respect to the original data.
There have recently been a number of approaches suggested for the integration of PH and deep learning, which we briefly review here.

In \cite{Chen2018b} a classification task was considered, and PH was used to regularise the decision boundary.
Typical regularisation of a decision boundary might encourage it to be smooth or to be far from the data.
Here, the boundary was encouraged to be simple from a topological point of view, meaning that topological complexities such as loops and handles in the decision boundary were discouraged.
\cite{Rieck2018} proposed a measure of the complexity of a neural network  using PH.
This measure of `neural persistence' was evaluated as a measure of structural complexity at each layer of the network, and was shown to increase during network training as well as being useful as a stopping criterion.

PH is applied to image segmentation, but the PH calculation has typically been applied to the input image and used as a way to generate features which can then be used by another algorithm.
Applications have included tumour segmentation \cite{Qaiser2016}, cell segmentation \cite{Assaf2017} and cardiac segmentation from computed tomography (CT) imaging \cite{Gao2013}. Recently \cite{Clough2019} proposed to use PH not to the input image being segmented, but rather to the candidate segmentation provided by the network. In an extended work Clough et al. \cite{Clough2020} the topological information found by the PH calculation can be used to provide a training signal to the network, allowing an differentiable loss function to compare the topological features present in a proposed segmentation, with those specified to exist by some prior knowledge.
\end{multicols}

%% file: tables/crf.tex
\footnotesize
\setlength\heavyrulewidth{0.25ex}
\setlength\lightrulewidth{0.25ex}
 \begin{tabularx}{\textwidth}{ S m L L}
    \caption{CNNs combined with CRF / MRF models}
    \label{tab:crf} \\ 
    \toprule
    \textbf{Authors} & \textbf{Anatomy} & \textbf{Title} & \textbf{Method} \\   
    \midrule
    \endhead 
    \multicolumn{4}{c}{CRF / MRF used for post-processing}\\
    \midrule
    \citet{journals/johe/zeju2017} & Brain Tumor & Low-Grade Glioma Segmentation Based on CNN with Fully Connected CRF & CRF refines CNN segmentation \\
    \hline
    \citet{DBLP:journals/neuroimage/WachingerRK18} & Brain Region & DeepNAT: Deep convolutional neural network for segmenting neuroanatomy & CRF refines hierarchical CNN segmentations \\
    \hline
    \citet{DBLP:journals/access/HuGZDXHCG19} & Brain Tumor & Brain Tumor Segmentation Using Multi-Cascaded Convolutional Neural Networks and Conditional Random Field & FC-CRF refines segmentations of three CNNs \\ 
    \hline
    \citet{shen2017} & Brain Tumor & Fully connected CRF with data-driven prior for multi-class brain tumor segmentation & Multiple FC-CRFs \\
    \hline    
    \citet{DBLP:journals/mia/KamnitsasLNSKMR17} & Brain Lesion & Efficient Multi-Scale 3D CNN with Fully Connected CRF for Accurate Brain Lesion Segmentation & FC-CRF refines two-pathway CNN \\
    \hline
    \citet{DBLP:conf/miccai/AlansaryKDKRMRH16} & Placenta & Fast Fully Automatic Segmentation of the Human Placenta from Motion Corrupted MRI & FC-CRF refines two-pathway CNN \\ 
    \hline
    \citet{DBLP:conf/isbi/ShakenTFLKPK16} & Sub-cortical regions & Sub-cortical brain structure segmentation using F-CNN's & MRF refines FCNN segmentation \\ 
    \hline
    \citet{DBLP:journals/jms/ZhaiL19} & Brain region & An Improved Full Convolutional Network Combined with Conditional Random Fields for Brain MR Image Segmentation Algorithm and its 3D Visualization Analysis & FC-CRF refines CNN with attention \\ 
    \hline
    \citet{DBLP:conf/miccai/Dou0JYQH16} & Liver & 3D Deeply Supervised Network for Automatic Liver Segmentation from CT Volumes & FC-CRF refines 3D FCNN with 3D supervision mechanism \\ 
    \hline
    \citet{DBLP:journals/mia/DouYCJYQH17} & Heart & 3D Deeply Supervised Network for Automated Segmentation of Volumetric Medical Images & FC-CRF refines cascading U-Nets \\ 
    \hline
    \citet{DBLP:conf/miccai/ChristEETBBRAHD16} & Liver & Automatic Liver and Lesion Segmentation in CT Using Cascaded Fully Convolutional Neural Networks and 3D Conditional Random Fields & FC-CRF refines cascaded FCNs \\ 
    \hline
    \citet{DBLP:conf/isbi/FuXWL16} & Retinal Vessel & Retinal vessel segmentation via deep learning network and fully-connected conditional random fields & FC-CRF refines CNN with side-outputs \\ 
    \hline
    \citet{DBLP:journals/titb/JinFWYLLZ18} & Left atrial appendage & Left Atrial Appendage Segmentation Using Fully Convolutional Neural Networks and Modified Three-Dimensional Conditional Random Fields & FC-CRF combines slices of FCN \\ 
    \hline
    \citet{DBLP:conf/miccai/CaiLXXY17} & Pancreas & Pancreas Segmentation in MRI using Graph-Based Decision Fusion on Convolutional Neural Networks & CRF refines results from FCN and HED network \\ 
    \hline
    \citet{DBLP:journals/jms/XiaYZ19} & Kidney & Deep Semantic Segmentation of Kidney and Space-Occupying Lesion Area Based on SCNN and ResNet Models Combined with SIFT-Flow Algorithm & MRF refines combined ResNet and SCNN \\ 
    \hline
    \citet{DBLP:journals/tmi/RajchlLOKPBDRHK17} & Fetal Brain / Lung & DeepCut: Object Segmentation from Bounding Box Annotations using Convolutional Neural Networks & Iterative CRF and CNN \\ 
    \hline
    \citet{DBLP:conf/miccai/NoguesLWRBLSTS16} & Lymph Node & Automatic Lymph Node Cluster Segmentation Using Holistically-Nested Neural Networks and Structured Optimization in CT Images & CRF refines HNN (FCN + DSN) segmentations \\ 
    \hline
    \citet{DBLP:conf/micad/YaguchiATO19} & Lung Nodules & 3D fully convolutional network-based segmentation of lung nodules in CT images with a clinically inspired data synthesis method & CRF refines 3D FCN segmentations \\ 
    \hline
    \citet{DBLP:conf/isbi/GaoXLWNSM16} & Lung & Segmentation label propagation using deep convolutional neural networks and dense conditional random field & CRF refines CNN segmentations \\ 
    \hline
    \citet{DBLP:journals/access/FengGQ20} & Brain Tumor & Study on MRI Medical Image Segmentation Technology Based on CNN-CRF Model & CRF refines DCNN segmentations \\ 
    \hline
    \citet{DBLP:journals/access/LiuZSLZG18} & Cervical Nuclei & Automatic segmentation of cervical nuclei based on deep learning and a conditional random field & Locally FC-CRF refines Mask-RCNN segmentation \\ 
    \hline
    \citet{DBLP:conf/icmip/ShenDLCQ18} & Brain Tumor & Brain tumor segmentation using concurrent fully convolutional networks and conditional random fields & Concurrent FCN refined by FC-CRF \\ 
    \hline
    \citet{DBLP:conf/icb/MesbahMM17} & Eye Sclera & Conditional random fields incorporate convolutional neural networks for human eye sclera semantic segmentation & Initial CNN boundaries refined by CRF \\ 
    \hline
    \citet{DBLP:conf/icpr/LuoY18} & Melanoma & Fast skin lesion segmentation via fully convolutional network with residual architecture and CRF & CRF refines FCN segmentations  \\ 
    \hline
    \citet{8986563} & Fundus Optic Disk & Improving the Performance of Convolutional Neural Network for the Segmentation of Optic Disc in Fundus Images Using Attention Gates and Conditional Random Fields & FC-CRF refines CNN segmentations \\ 
    \hline
    \citet{DBLP:journals/access/QiuCQZ20} & Skin Lesion & Inferring Skin Lesion Segmentation With Fully Connected CRFs Based on Multiple Deep Convolutional Neural Networks & CRF refines segmentations of DCNN ensemble \\ 
    \hline
    \citet{DBLP:conf/miccai/NguyenPMSPHWCS18} & Ocular structures & Ocular structures segmentation from multi-sequences mri using 3d unet with fully connected crfs & FC-CRF refines CNN segmentations \\ 
    \hline
    \citet{DBLP:conf/isbi/CaoZSBMERS19} & Prostate cancer lesions & Prostate Cancer Detection and Segmentation in Multi-parametric MRI via CNN and Conditional Random Field & Selective Dense CRF refines CNN segmentations \\
    \midrule
    \multicolumn{4}{c}{CNN and CRF trained end-to-end} \\
    \midrule
    \citet{DBLP:journals/mia/ZhaoWSLZF18} & Brain Tumor & A deep learning model integrating FCNNs and CRFs for brain tumor segmentation. & Combination of FCNN and CRF-RNN \\
    \hline
    \citet{DBLP:journals/corr/abs-1807-07464} & Prostate / Brain Tumor & Conditional Random Fields as Recurrent Neural Networks for 3D Medical Imaging Segmentation & Combination of FCNN and CRF-RNN \\ 
    \hline
    \citet{DBLP:conf/miccai/FuXLW016} & Retinal Vessel & DeepVessel: Retinal Vessel Segmentation via Deep Learning and Conditional Random Field & Combination of CNN and CRF-RNN layers \\ 
    \hline
    \citet{DBLP:journals/corr/abs-1811-03549} & White matter hyperintensities & An End-to-end Approach to Semantic Segmentation with 3D CNN and Posterior-CRF in Medical Images & Combination of U-Net and FC-CRF \\ 
    \hline
    \citet{DBLP:journals/cars/XuZL18} & Bladder & Automatic bladder segmentation from CT images using deep CNN and 3D fully connected CRF-RNN & Combination of CNN and CRF-RNN \\ 
    \hline
    \citet{DBLP:journals/access/DengSWZN20} & Brain Tumor & Deep Learning-Based HCNN and CRF-RRNN Model for Brain Tumor Segmentation & Combination of HCNN and CRF-RRNN \\ 
    \hline
    \citet{DBLP:journals/corr/abs-2008-04488} & Prostate & ARPM-net: A novel CNN-based adversarial method with Markov Random Field enhancement for prostate and organs at risk segmentation in pelvic CT images & CNN combined with MRF block \\ 
    \hline
    \citet{DBLP:conf/miccai/ZhaoWSLFZ16} & Brain tumor & Brain tumor segmentation using a fully convolutional neural network with conditional random fields & CRF integrated into FCNN \\ 
    \hline
    \citet{10.1007/978-981-10-5230-9_17} & Retinal Vessel & Efficient CNN-CRF network for retinal image segmentation & Combination of CNN and CRF \\  \bottomrule
\end{tabularx}
\normalsize

%% file: tables/asm.tex
\footnotesize
\setlength\heavyrulewidth{0.25ex}
\setlength\lightrulewidth{0.25ex}
\begin{tabularx}{\textwidth}{ S m L L}
    \caption{CNNs combined with Active Shape Models}
    \label{tab:asm} 
    \\ \toprule
    \textbf{Authors} & \textbf{Anatomy} & \textbf{Title} & \textbf{Method} \\
    \midrule
    \endhead 
    \multicolumn{4}{c}{ASM for post-processing}\\
    \midrule
    \citet{DBLP:journals/tmi/XingXY16} & Nucleus & An Automatic Learning-Based Framework for Robust Nucleus Segmentation & Shape Model refines CNN segmentation
    \\ \hline
    \citet{DBLP:journals/access/HeXHJ18} & Prostate & Automatic Magnetic Resonance Image Prostate Segmentation Based on Adaptive Feature Learning Probability Boosting Tree Initialization and CNN-ASM Refinement & Three-level-ASM refines segmentations of CNN
    \\ \hline
    \citet{DBLP:journals/cars/FauserSBHKKSM19} & Temporal Bone & Toward an automatic preoperative pipeline for image-guided temporal bone surgery & Probabilistic ASM refines 2D U-Net segmentation
     \\ \hline
    \citet{DBLP:journals/tmi/LiHTCST18} & Myocardial & Fully Automatic Myocardial Segmentation of Contrast Echocardiography Sequence Using Random Forests Guided by Shape Model & ASM refines random-forest segmentations initialized by a CNN
     \\ \hline
     \citet{DBLP:journals/tip/MedleySN20} & Left Ventricle & Deep Active Shape Model for Robust Object Fitting & ASM initialized with CNN generated features maps
     \\ \hline
     \citet{DBLP:journals/mia/KarimiZMAMSAS19} & Prostate & Accurate and robust deep learning-based segmentation of the prostate clinical target volume in ultrasound images & SSM refines segmentations from ensemble of CNNs
     \\ \hline
     \citet{DBLP:conf/isbi/TabriziMCJL18} & Kidney & Automatic kidney segmentation in 3D pediatric ultrasound images using deep neural networks and weighted fuzzy active shape model & Fuzzy ASM segmentations based on DNN generated bounding boxes
     \\ \hline
     \citet{DBLP:journals/access/Hsu19} & Left Ventricle & Automatic Left Ventricle Recognition, Segmentation and Tracking in Cardiac Ultrasound Image Sequences & ASM improves R-CNN segmentations for detection and tracking
     \\ 
     
    \midrule
    \multicolumn{4}{c}{ASM as prior-knowledge}\\
    \midrule
    \citet{DBLP:journals/corr/abs-1802-01268} & Brain Region & Accurate brain extraction using Active Shape Model and Convolutional Neural Networks & CNN refines ASM segmentations
    \\ \hline
    \citet{DBLP:conf/miip/ChengRLWTGMACSM16} & Prostate & Active appearance model and deep learning for more accurate prostate segmentation on MRI & 2D-CNN refines segmentations from an Active Appearance Model
     \\ \hline
     \citet{DBLP:conf/miip/FanZWND20} & Intra-Cholear Anatomy & Combining model- and deep-learning-based methods for the accurate and robust segmentation of the intra-cochlear anatomy in clinical head CT images & U-Net refines ASM segmentations
     \\ \hline
     \citet{DBLP:conf/midl/NguyenPHWRSSC19} & Uveal Melanoma & A novel segmentation framework for uveal melanoma based on magnetic resonance imaging and class activation maps & U-Net segmentations based on a CRF that uses ASM as prior knowledge
     \\ 
     
    \midrule
    \multicolumn{4}{c}{Pipelines with multiple ASM and CNN models \& Hybrid approaches}\\
    \midrule
    \citet{DBLP:journals/mia/AmbellanTEZ19} & Knee Bone / Cartilage & Automated Segmentation of Knee Bone and Cartilage combining Statistical Shape Knowledge and Convolutional Neural Networks: Data from the Osteoarthritis Initiative & Three CNN and two SSM models
    \\ \hline
    \citet{jorunal/oac/tack2018} & Knee Menisci & Knee Menisci Segmentation using Convolutional Neural Networks: Data from the Osteoarthritis Initiative & 3D CNN and SSM initialized by 2D models
    \\ \hline
    \citet{10.3389/fnins.2020.00015} & Hippocampus & Shape Information Improves the Cross-Cohort Performance of Deep Learning-Based Segmentation of the Hippocampus & ASM as input for CNN
    \\ \hline
    \citet{DBLP:conf/miccai/MaLWE18} & Pancreas & A novel bayesian model incorporating deep neural network and statistical shape model for pancreas segmentation & U-Net and SSM segmentations combined within Bayesian model
    \\ \hline
    \citet{DBLP:journals/corr/abs-2007-11726} & Prostate & A weakly supervised registration-based framework for prostate segmentation via the combination of statistical shape model and CNN & Segmentations combined of U-Net and SSM-Net predictions
    \\ \hline
    \citet{DBLP:journals/corr/abs-2010-08952} & Left Ventricle & Shape Constrained CNN for Cardiac MR Segmentation with Simultaneous Prediction of Shape and Pose Parameters & Hybrid approach where CNN generates segmentations and ASM parameters
    \\ \hline
    \citet{DBLP:journals/cars/KarimiSKNS18} & Prostate & Prostate segmentation in MRI using a convolutional neural network architecture and training strategy based on statistical shape models & CNN predicts segmentations and 3D-ASM parameters 
    \\ \hline
    \citet{DBLP:conf/miccai/SchockKAHKTCAKN20} & Knee Bone \& Cartilage & A Method for Semantic Knee Bone and Cartilage Segmentation with Deep 3D Shape Fitting Using Data from the Osteoarthritis Initiative & CNN that predicts segmentations and 3D-ASM parameters is refined by U-Net
    \\
    \bottomrule
\end{tabularx}
\normalsize

%% file: tables/contours.tex
\footnotesize
\setlength\heavyrulewidth{0.25ex}
\setlength\lightrulewidth{0.25ex}
\begin{tabularx}{\textwidth}{ S m L L}
    \caption{CNNs combined with Active Contour Models}
    \label{tab:acm} 
   \\ \toprule
   \textbf{Authors} & \textbf{Anatomy} & \textbf{Title} & \textbf{Method}
   \\ \midrule
    \endhead 
    \multicolumn{4}{c}{ACM for post-processing}\\
    \midrule
    \citet{jorunal/mep/middelton2004} & Lung & Segmentation of magnetic resonance images using a combination of neural networks and active contour models & ACM refines MLP segmentation
    \\ \hline
    \citet{DBLP:journals/sivp/SalimiPM18} & Prostate & Fully automatic prostate segmentation in MR images using a new hybrid active contour-based approach & ACM refines MLP segmentation
     \\ \hline
    \citet{DBLP:conf/bibm/LiLYW17} & Left Ventricle & Left ventricle segmentation by combining convolution neural network with active contour model and tensor voting in short-axis MRI & ACM refines FCN segmentation
     \\ \hline
    \citet{journal/mp/hu2018} & Breast Tumor & Automatic tumor segmentation in breast ultrasound images using a dilated fully convolutional network combined with an active contour model & Phase-based ACM refines dilated FCN segmentation
    \\ \hline
    \citet{jorunal/mp/guo2019} & Liver & Automatic liver segmentation by integrating fully convolutional networks into active contour models & ACM refines multi-branch FCN segmentation
    \\ \hline
    \citet{DBLP:conf/iconip/ZhaoWFQ18} & Nucleus & Improved Nuclear Segmentation on Histopathology Images Using a Combination of Deep Learning and Active Contour Model & Hybrid ACM refines multi-branch FCN segmentation
    \\ \hline
    \citet{DBLP:journals/corr/abs-1908-06933} & Liver / Brain Lesion / Lung & Deep Active Lesion Segmentation & ACM refines signed distance maps from FC-CNN
 \\ \hline
    \citet{DBLP:conf/miccai/TangVZCJ17} & Liver / Left Ventricle & A Deep Level Set Method for Image Segmentation & Level-set ACM refines FCN segmentations iteratively
     \\ \hline
    \citet{journal/mp/chu2016} & Bladder & Urinary bladder segmentation in CT urography using deep-learning convolutional neural network and level sets & Multiple level-set functions segment CNN output ROIs
 \\ \hline
    \citet{DBLP:journals/tmi/HoogiSVR17} & Liver Lesion & Adaptive Estimation of Active Contour Parameters Using Convolutional Neural Networks and Texture Analysis & Level-set function iteratively improves CNN segmentation
 \\ \hline
    \citet{joruanl/jomiahi/fang2019} & Breast Tumor & Combining a Fully Convolutional Network and an Active Contour Model for Automatic 2D Breast Tumor Segmentation from Ultrasound Images & Phase-based ACM refines initial contours from dilated FCNN
    \\ \hline
    \citet{journal/jomi/xu2019} & Breast Cancer Nuclei & Convolutional neural network initialized active contour model with adaptive ellipse fitting for nuclear segmentation on breast histopathological images & ACM refines CNN segmentations
    \\ \hline
    \citet{DBLP:conf/miip/MaY19} & Teeth & Automatic dental root CBCT image segmentation based on CNN and level set method & ACM refines CNN segmentations
    \\ \hline
    \citet{10.1117/12.2556928} & Ventricles & Active contours for multi-region segmentation with a convolutional neural network initialization & Phase level-set function refines CNN segmentations
    \\ \hline
    \citet{DBLP:conf/cinc/0003WLZ19} & Left Atrium & A Framework for Left Atrium Segmentation on CT Images with Combined Detection Network and Level Set Model & 3D level-set model initialized by Faster RCNN
    \\ \hline
    \citet{YANG2021108} & Teeth & Accurate and automatic tooth image segmentation model with deep convolutional neural networks and level set method & Level-set based on contours derived from U-Net predictions
    \\ \hline
    \citet{DBLP:conf/ijcnn/NunesMSBF20} & Lung & Adaptive Level Set with region analysis via Mask R-CNN: A comparison against classical methods & ACM improves Mask R-CNN segmentations  
    \\ \hline
    \citet{DBLP:journals/cbm/XieSC20} & Left Ventricle & Automatic left ventricle segmentation in short-axis MRI using deep convolutional neural networks and central-line guided level set approach & Level-set model improves CNN initialization
    \\ \hline
    \citet{DBLP:conf/isicdm/GongZZZG19} & Pancreas & Convolutional Neural Networks Based Level Set Framework for Pancreas Segmentation from CT Images & Level-set model based on initial contour from CNN
    \\ \hline
    \citet{DBLP:conf/miccai/ZhangDL20} & Cervical Cell / Skin Lesion & Deep Active Contour Network for Medical Image Segmentation & ACM integrated into CNN that learns initial parameters (end-to-end)
    \\ \hline
    \citet{DBLP:journals/sivp/ZhangWLW20} & Plaque & Faster R-CNN, fourth-order partial differential equation and global-local active contour model (FPDE-GLACM) for plaque segmentation in IV-OCT image & ACM initialized with bounding box from R-CNN
    \\ \hline
    \citet{DBLP:journals/mbec/SilvaDFFSPC20} & Prostate & Superpixel-based deep convolutional neural networks and active contour model for automatic prostate segmentation on 3D MRI scans & ACM refines DCNN segmentations
    \\ \hline
    \citet{DBLP:conf/ijcnn/KotKSC20} & Brain Tumor & U-Net and Active Contour Methods for Brain Tumour Segmentation and Visualization & ACM refines U-Net segmentations
    \\ \hline
    \citet{DBLP:journals/mia/AvendiKJ16} & Left Ventricle & A combined deep-learning and deformable-model approach to fully automatic segmentation of the left ventricle in cardiac MRI & CNN and AE initialize level set function
    \\

    \midrule
    \multicolumn{4}{c}{CNN refines ACM}\\
    \midrule   
    \citet{DBLP:journals/eswa/KasinathanJGRFP19} & Lung Tumor / Nodule & Automated 3-D Lung Tumor Detection and Classification by an Active Contour Model and CNN Classifier & CNN refines multiple ACM segmentations
    \\ \hline
    \citet{DBLP:journals/corr/RupprechtHBN16} & Left ventricular cavity & Deep Active Contour & CNN refines ACM
    \\ \hline 
    \citet{DBLP:conf/socpar/AhmedMK09} & Brain & A Hybrid Approach for Segmenting and Validating T1-Weighted Normal Brain MR Images by Employing ACM and ANN & ANN based on ACM preprocessed images
 \\ 
    \bottomrule
\end{tabularx}
\normalsize

%% file: sections/discussion.tex
\section{Discussion} 
\label{Discussion}
As the deep learning research effort for medical image segmentation is consolidating towards incorporating shape constraints to ensure downstream analysis, certain patterns are emerging as well. In the next few subsections, we discuss such clear patterns and emerging questions relevant for the progress of research in this direction.

\subsection{End-to-End vs post/pre-hoc} With the maturity of research, this field is clearly moving beyond post-/pre-hoc setting towards more systematic end-to-end training approaches. This effect is depicted in Figure \ref{e2e_shift} where the paper counts are aggregated from this work and \cite{DBLP:journals/corr/abs-2011-08018}. The maturity of deep learning frameworks (especially PyTorch), novel architectures (especially generative modeling) and automatic differentiation make it possible to incorporate  complex shape-based loss functions during training. With the availability of these tools, large models can be trained with tailored shape streams in the model architecture to incorporate shape information.

\begin{Figure}
    \centering
    \includegraphics[scale=.55]{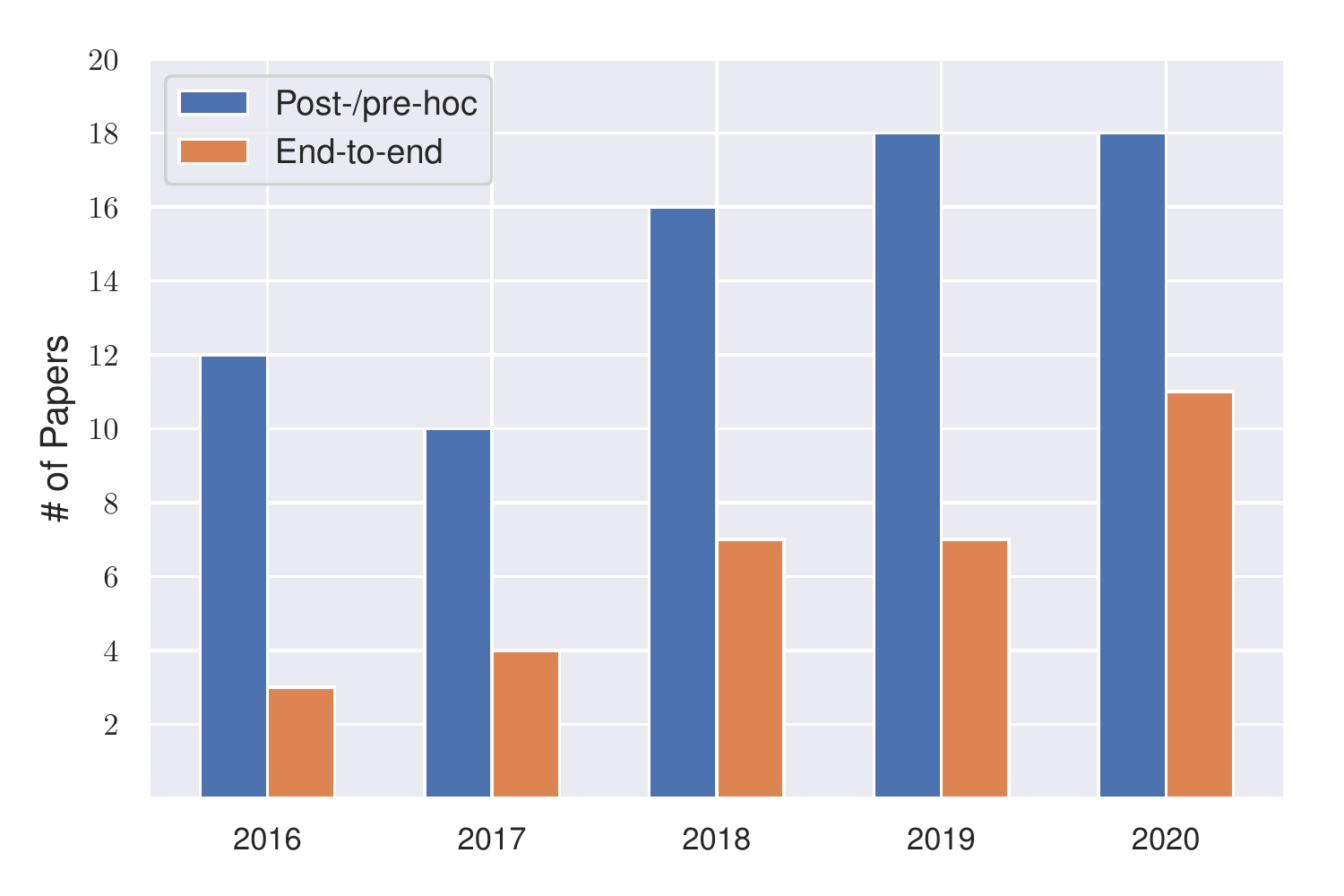}
	\captionof{figure}{Temporal trend towards end-to-end approaches}
	\label{e2e_shift}
\end{Figure}

\subsection{Semi-supervised segmentation} The ability to incorporate additional information using shape as a prior can aid in reducing the total number of necessary annotations in achieving a good segmentation. The shape priors can useful in generating controlled data augmentations for the medical image analysis task in hand and reduce the number of unrealistic augmentations. This would be instrumental in particular in the case of rare diseases, where there is not enough of data and manual annotations to train a neural network. The shape priors that are giving clues about the expected pathology in such cases can lead to better segmentation accuracy in the final output.

\subsection{Effectiveness in pathological cases} One common theme identified by last few decades worth research on shape modeling is the difficulty in representing the pathological shapes. While the "typical shapes" i.e. normal shapes lie in a low-dimensional sub-manifold, the pathological cases have a long tail in the distribution (e.g. congenital heart diseases). That is normal shapes are self-similar but pathological cases contain atypical shapes along with typical pathologies. Traditional linearized shape modeling had trouble addressing this issue whereas the non-linear modeling of shape statistics had its issue in terms of intractable numerics. Whether a neural approach can address this overarching problem of encoding pathological shapes is an open problem. Unfortunately, from our literature search, we have not found any clear direction to address this perennial issue of shape modeling.

\subsection{Evaluation} While the shape constraints are becoming increasingly commonplace for medical image segmentation, we believe the visual perception and human comprehension plays a significant role behind the interest of the community. The more general question of real world effectiveness of these methods are not often studied. For example, how effective these shape constraints are under noisy annotation is an open question? While the segmentation quality is most often measured by the Dice metric, \cite{DBLP:journals/corr/abs-1806-02051} has already prescribed to move beyond Dice to evaluate the segmentation quality. Topological accuracy of anatomical structures is increasingly used as an evaluation metric to address the shortcomings of classical image segmentation evaluation metric in medical image analysis \cite{byrne2020persistent}. Finally, segmentation is typically a mean to an end. As such, the effectiveness of these segmentation techniques should be measured quantitatively for downstream evaluation tasks such as visualization, planning~\cite{DBLP:journals/cars/FauserSBHKKSM19} etc.

%% file: sections/conclusion.tex
\section{Conclusion}
Bringing prior knowledge about the shape of the anatomy for semantic segmentation is a rather well-trodden idea.
The community is devising new ways to incorporate such prior knowledge in deep learning models trained with frequentist approach.
While the Bayesian interpretation of deep learning segmentation networks is an upcoming trend, it is already shown that under careful considerations, prior knowledge about the shape can be incorporated even in frequentist approaches with significant success. 

We see the future research concentrating more on end-to-end networks with the overarching theme of learning using Analysis-by-synthesis. 
Early work has demonstrated the effectiveness of shape constraints in federated learning and this will be a major direction in the coming years. 

We believe the community needs to address the issues discussed in Section~\ref{Discussion} before shape constrained segmentation can be considered as a trustworthy technology in practical medical image analysis.
To this end, we can think of shape constrained segmentation as a technical building block within a bigger image analysis pipeline rather than a stand-alone piece of technology.
For example, in the case of surgical planning and navigation pipeline, such shape constraints can be meaningful provided the performance is thoroughly validated under pathological cases with multiple quality metrics. Important steps have already been taken in this direction. 
In short, along with exciting results, shape constrained deep learning for segmentation opens up many possible research questions for the next few years. 
Proper understanding and answering those hold the key to their successful deployment in the real clinical scenario.